\documentclass[a4paper,twosides]{article}
\usepackage{fontspec}
\usepackage{float}
\usepackage{units}

\usepackage{multicol,multirow,graphicx,fancyhdr,epstopdf}
\usepackage{amsmath,amsfonts,amssymb,bm,upgreek,mathrsfs,ccmap,mathcomp}
\usepackage[pagewise,switch,columnwise]{lineno}
\usepackage[compress,nospace]{cite}
\usepackage[dvipsnames]{xcolor}
\usepackage{CPL-2023}
\usepackage{longtable}
\usepackage{capt-of}

\newcommand{\cplyear}{2026} \newcommand{\cplvol}{xx}
 \newcommand{\cplpagenumber}{xxxxxx}
 \newcommand{\cplpage}{\cplpagenumber-\thepage}

\begin{document}

\twocolumn[

\vspace*{-4mm}
\begin{center}

\par\vspace{3mm}
\large\bf{\boldmath{From Detection to Host Galaxy Identification: Precision Continuous Gravitational Wave Localization with a Few Anchor Pulsars}}

\par\vspace{3mm}

\normalsize \rm{}
Chao-Fan Wen $^{1,2}$, %
Yi-Qin Chen $^{1,2}$, %
Shi-Yi Zhao $^{1,2*}$, %
Hao Ding $^{3,4*}$, %
Xingjiang Zhu $^{1*}$, %

\par\vspace{3mm}

$^{1}$Institute for Frontier in Astronomy and Astrophysics \& Faculty of Arts and Sciences, Beijing Normal University, Zhuhai 519087, China

$^{2}$School of Physics and Astronomy, Beijing Normal University, Beijing 100875, China

$^{3}$National Astronomical Observatories, Chinese Academy of Sciences, Beijing 100012, China

$^{4}$Korea Astronomy and Space Science Institute, 776 Daedeokdae-ro, Yuseong-gu, Daejeon, Republic of Korea

\end{center}
\vskip 1.5mm

\small{\narrower Pulsar Timing Arrays (PTAs) are rapidly advancing toward the detection of continuous gravitational waves from individual supermassive binary black holes.
While it is well established that coherently utilizing the ``pulsar term" requires astrometric distance uncertainties to be smaller than the gravitational wavelength, achieving this precision across an entire array is observationally prohibitive.
Here, we demonstrate that achieving sub-wavelength precision for a few ``anchor" pulsars is sufficient to phase-lock the array and drastically shrink the sky-localization error. Using 20 years of realistically simulated data, we systematically evaluate the localization performance of a 25-pulsar array containing three to six high-precision anchors. We show that while introducing three sub-wavelength anchors can reduce the 90\% credible sky area by a factor of 30 in certain directions, expanding this high-precision subset to six anchor pulsars ensures high-precision localizations across diverse source directions.
Evaluating a representative set of sky directions, including local galaxy clusters and the locations of maximum and minimum array sensitivity, this six-anchor configuration yields 90\% credible localization areas ranging from $\sim 0.1$ to $9.2 \text{ deg}^2$ at a signal-to-noise ratio of 20. Furthermore, once this minimal subset crosses the sub-wavelength threshold, further reductions in distance uncertainty yield diminishing returns. This establishes a highly efficient near-term observational strategy: prioritizing intensive parallax campaigns for a small core of stable millisecond pulsars provides a cost-effective pathway to precision multi-messenger astronomy.
\par}\vskip 3mm


]

\footnotetext{\hspace*{-5.4mm}$^{*}$Corresponding authors. 
Email: zhuxj@bnu.edu.cn; 

hdingastro@hotmail.com; zhaosyastro@gmail.com

\noindent\copyright\,{\cplyear}
\href{http://www.cps-net.org.cn}{Chinese Physical Society} and
\href{http://www.iop.org}{IOP Publishing Ltd}}

\textbf{Introduction.}
PTAs are opening a new observational window into the Universe, having recently found strong evidence for nanohertz-frequency gravitational waves (GWs). This signal currently manifests as a stochastic gravitational-wave background \ucite{agazie2023nanograv, antoniadis2023second3, reardon2023search, xu2023searching, miles2025meerkat}, which is likely driven by an ensemble of inspiraling supermassive binary black holes (SMBBHs). As observational baselines lengthen and timing precision improves with next-generation radio facilities, the field is rapidly transitioning from characterizing this unresolved background toward the identification of individual continuous GW (CGW) sources \ucite{rosado2015expected, burke2019astrophysics, falxa2023searching, agazie2023nanogravsmbbh, antoniadis2024second, zhao2025searching}. Localizing these individual SMBBHs is a critical next step for multi-messenger astronomy, as identifying an electromagnetic counterpart would yield 
unprecedented insights into galaxy merger histories, accretion dynamics, and the ``final parsec problem"—the theoretical puzzle of how these massive binaries shed their remaining orbital energy to shrink their separation below the parsec scale and eventually merge \ucite{yu2002evolution, milosavljevic2003final, khan2013supermassive, kormendy2013coevolution, de2019quest, taylor2021nanohertz}.

However, precise sky localization of a CGW source using a PTA remains a profound observational challenge. The timing residual induced by a single CGW consists of two components: the `Earth term', which captures the local spacetime strain at the solar system, and the `pulsar term', which encodes the strain at the pulsar thousands of years earlier \ucite{zhu2016detection, wang2017pulsar,chen2022parameter}. Standard PTA searches rely predominantly on the Earth term to reconstruct the source parameters.
To fully harness the coherent power of the pulsar term in GW data analysis, one must tightly constrain the Earth-pulsar distance to sub-wavelength precision. For the nanohertz GWs targeted by PTAs, this wavelength ($\lambda_{\rm GW}$) is on the order of parsecs (e.g., 2.4 pc for $f_{\rm GW}= 4~{\rm nHz}$).
Typical pulsar distances are of order $\sim 1\text{ kpc}$.
Because most pulsar distance estimates currently suffer from fractional uncertainties $\gtrsim 20\%$, the distance error drastically exceeds the gravitational wavelength ($D_{\rm err} \gg \lambda_{\rm GW}$).
Consequently, the phase of the pulsar term is effectively scrambled, breaking phase coherence across the array and severely degrading the network's spatial resolving power.

Overcoming this localization bottleneck requires independent high-precision astrometry. Specifically, one must resolve sub-parsec distances for pulsars that are typically located several kiloparsecs from Earth. Currently, this pushes the absolute limits of astrometric capabilities; only a couple of millisecond pulsars have distance measurements approaching the required parsec-level precision. Looking ahead, advancements in Very Long Baseline Interferometry (VLBI) and the impending full operation of the Square Kilometre Array (SKA) promise to dramatically refine pulsar parallax measurements\ucite{shamohammadi2024meerkat, ding2025systemic}. Even with these future facilities, securing sub-wavelength distance precision for an entire PTA consisting of dozens of pulsars will remain practically infeasible in the foreseeable future. However, achieving this high precision for a sub-array of three to six pulsars may soon be within reach.

Previous studies have highlighted the profound benefits of constraining the pulsar term. Analytical analyses suggested that pulsars with precisely known distances act analogously to baselines in a diffraction-limited interferometer, significantly enhancing angular resolution\ucite{BoylePen12, tsai2025reaching}. More recently, simulations demonstrated that achieving sub-wavelength distance precision for two nearby pulsars (J0437$-$4715, J0030+0451) can reduce source localization uncertainties by up to several orders of magnitude\ucite{kato2023precision, kato2026realistic}. However, these estimates relied on analytical posterior approximations, specifically assuming a high signal-to-noise ratio limit, and idealized datasets limited exclusively to Gaussian white noise. Such idealizations do not fully capture the complex realities of actual PTA observations, which must contend with significant red noise, irregular observational cadences, and mixed-array configurations where distance uncertainties vary drastically between pulsars.

To address these limitations, this paper quantitatively evaluates the impact of high-precision pulsar distances on CGW source localization under realistic observational conditions. Utilizing 20 years of simulated PTA data that rigorously incorporates empirical white noise, red noise, and actual observational cadences, we inject a CGW signal and recover its parameters using Bayesian inference. We explore a realistic, mixed-array scenario where only a small subset of ``anchor" pulsars achieves high-precision distance priors, while the remaining pulsars retain typical 20\% distance errors. 
The remainder of this paper is organized as follows. We first describe the simulations and methodology, then present the parameter-estimation results, discuss the observational feasibility of the proposed astrometric thresholds and their practical implications for future multi-messenger astronomy, and finally summarize our main conclusions.

\medskip

\textbf{Simulations and Methodology}

\textit{1. Pulsar Selection and Noise Modeling.}
To ensure a realistic PTA configuration, we construct our pulsar sample using publicly available datasets from the Parkes Pulsar Timing Array third data release (PPTA DR3)\ucite{Zic2023PPTADR3}, the European Pulsar Timing Array second data release (EPTA DR2)\ucite{antoniadis2023second1}, the NANOGrav 15-yr data set\ucite{Agazie2023}, and the MeerKAT 4.5-yr data set\ucite{Miles2025MeerKATdata}. For each pulsar, we adopt the reported white- and red-noise parameters as well as the measured timing uncertainties.

We first select six pulsars with precisely measured distances, including five nearby ($\leq0.4$\,kpc) pulsars---J0030$+$0451, J0437$-$4715, J0636$-$3044, J1744$-$1134 and J2222$-$0137---and arguably the best-timed pulsar J1909$-$3744.
These pulsars are selected independently of their timing performance and are included in all search strategies considered below as a core subset.
To investigate a scenario where fewer pulsars achieve sub-parsec distance determinations in future CGW searches, we also define an even smaller sub-array of three pulsars (J0030$+$0451, J0437$-$4715, and J2222$-$0137).

While only a small number of the six selected pulsars currently have distance measurements at the parsec level, upcoming high-sensitivity timing and VLBI campaigns are expected to further improve these constraints. Specifically, reaching a parallax precision of $\lesssim 10~\mu\text{as}$ will serve as the critical threshold for unlocking sub-parsec astrometry for other nearby pulsars. Although we have carefully grounded our analysis in a realistic PTA sample, actual astrometric precision depends on a complex suite of observational factors.
For simplicity, we assume a distance uncertainty of 1 pc for all six pulsars in our analysis.
Forecasting the exact astrometric capabilities of future facilities, or predicting the definitive list of pulsars that will achieve this precision, is beyond the scope of this paper.

We emphasize that our choice to cap the high-precision sub-array at six pulsars is dictated by observational reality rather than computational scope. Achieving the sub-parsec distance precision necessary to phase-lock the array requires extraordinary parallax measurements. The six pulsars selected for our core subset represent the nearest and best-timed millisecond pulsars available. Beyond this specific group, it is unlikely that the required parallax precision can be achieved within a reasonable $\lesssim10$-year timeframe, even with the advent of SKA-mid VLBI or ngVLA.

While the underlying public PTA data releases exhibit significant heterogeneity in observational cadence, timing methodology, and chromatic-noise treatment, our simulations adopt a unified pulsar-level description. To handle overlapping pulsars across different arrays, we eliminate double-counting by retaining only the single representation that yields the lowest empirical weighted root-mean-square (WRMS) timing residual. Furthermore, to maintain a controlled baseline for our GW parameter estimation, we intentionally bypass dataset-specific systematic, such as DMX versus DM Gaussian-process modeling or narrowband/wideband distinctions, assuming these observational complexities have been absorbed into our empirical noise budget. Our aim is not to reconstruct each PTA data set in full detail, but rather to build a observationally motivated sample suitable for studying the impact of high-precision pulsar distances.

Using the observed noise parameters, including the EFAC and EQUAD white-noise scaling parameters, and modeling red noise as a stationary Gaussian process with a power-law power spectral density,
we generate 20-year simulated timing datasets for each pulsar. To evaluate timing performance and optimize GW sensitivity, we compute the WRMS of the simulated timing residuals.
Excluding the six core pulsars, we rank the remaining pulsars according to their WRMS values. We then select 19 pulsars with the lowest WRMS and combine them with the core subset, leading to a PTA consisting of 25 pulsars. The adopted noise parameters and pulsar distances are summarized in Appendix Table \ref{tab:pulsar_params}.

\textit{2. Signal Model.}
The timing residuals for each pulsar are modeled as\ucite{arzoumanian2023nanograv}
\begin{equation}
r = M\boldsymbol{\epsilon} + n_{\rm white} + n_{\rm red} + s,
\label{eq:timing_residuals}
\end{equation}
where $M$ is the design matrix obtained from the linearized timing model and $\boldsymbol{\epsilon}$ denotes the vector of timing-model parameter offsets. The term $M\boldsymbol{\epsilon}$ represents the contribution from imperfect timing-model subtraction, $n_{\rm white}$ and $n_{\rm red}$ denote the white and red noise components, and $s$ is the GW signal contribution.

For a single non-spinning SMBBH in a circular orbit, the induced timing perturbation consists of two components: the ``Earth term" evaluated at time $t_e$, and the ``pulsar term" evaluated at time $t_p$. These times are related by the geometric light-travel time:
\begin{equation}
t_p = t_e - D_{p}(1 + \hat{\Omega} \cdot \hat{p}),
\end{equation}
where $D_{p}$ is the pulsar distance, $\hat{\Omega}$ is the unit vector pointing from the GW source to the Solar System barycenter, and $\hat{p}$ is the unit vector pointing toward the pulsar.

The full signal $s(t)$ is the antenna-pattern-weighted sum of the plus ($+$) and cross ($\times$) polarization modes, which for an evolving circular binary are expressed as:
\begin{equation}
\begin{aligned}
s_{+}(t) &= 
\frac{\mathcal{M}^{5/3}}{d_L \omega(t)^{1/3}}
\Big\{
-\sin[2\Phi(t)](1+\cos^2\iota)\cos 2\psi \\
&\qquad- 2\cos[2\Phi(t)]\cos\iota\sin 2\psi
\Big\}, \\
s_{\times}(t) &= 
\frac{\mathcal{M}^{5/3}}{d_L \omega(t)^{1/3}}
\Big\{
-\sin[2\Phi(t)](1+\cos^2\iota)\sin 2\psi \\
&\qquad+ 2\cos[2\Phi(t)]\cos\iota\cos 2\psi
\Big\},
\end{aligned}
\label{eq:signal}
\end{equation}
where $\mathcal{M} \equiv (m_1 m_2)^{3/5}/(m_1+m_2)^{1/5}$ is the chirp mass, $m_1$ and $m_2$ denote the masses of the two black holes, respectively, $d_L$ is the luminosity distance to the source, $\psi$ is the polarization angle, and $\iota$ is the binary inclination angle.
The orbital angular frequency and phase of Equation (\ref{eq:signal}) are given by
\begin{equation}
\omega(t) =
\omega_0
\left[
1 - \frac{256}{5}
\mathcal{M}^{5/3}
\omega_0^{8/3}
(t - t_0)
\right]^{-3/8},
\label{eq:freq_evolution}
\end{equation}
\begin{equation}
\Phi(t) =
\Phi_0 +
\frac{1}{32}
\mathcal{M}^{-5/3}
\left[
\omega_0^{-5/3} - \omega(t)^{-5/3}
\right],
\label{eq:phase_evolution}
\end{equation}
where $t_0$ is taken to be the epoch of the earliest observation in the dataset, $\omega_0 \equiv \omega(t_0) = \pi f_{\rm GW}$ is the initial orbital angular frequency, and $\Phi_0$ is the initial orbital phase of the Earth term.

The CGW signal model is therefore characterized by the parameter set
\[
\{\theta, \phi, \psi, \iota, \Phi_0, f_{\rm GW}, d_L, \mathcal{M}\},
\]
where $(\theta,\phi)$ specify the sky location of the source; here $\phi=\alpha$ and $\cos\theta=\sin\delta$, where $\alpha$ and $\delta$ are the right ascension (RA) and declination (Dec), respectively.

\textit{3. Simulated Data.}
The optimal signal-to-noise ratio for a CGW signal in a PTA is given by
\begin{equation}
(S/N)^{2} = \sum_{i=1}^{N_{p}} 
\mathbf{S}_{i}^{T} 
\mathbf{G}_{i}
\left(\mathbf{G}_{i}^{T} \mathbf{C}_{i} \mathbf{G}_{i}\right)^{-1} 
\mathbf{G}_{i}^{T} 
\mathbf{S}_{i},
\label{eq:snr}
\end{equation}
where $N_p$ is the number of pulsars, 
$\mathbf{S}_i$ denotes the CGW-induced timing residuals, 
$\mathbf{C}_i$ is the noise covariance matrix, 
and $\mathbf{G}_i$ projects the data into the timing-model–marginalized subspace.

To establish a controlled baseline for our simulations, all intrinsic CGW parameters are kept the same except luminosity distance. We perform a full-sky scan with the 25-pulsar array to determine the detection horizon. We define our reference sky location at $(\theta,\phi)=(2.60,1.08)$ rad, which corresponds to the direction yielding the maximum distance for $S/N =20$. Unless otherwise specified, all primary simulations adopt this sky location, and we scale the GW source luminosity distance to ensure $S/N =20$ across different PTA/source configurations.

We assume a fiducial circular binary with a chirp mass of $\mathcal{M}=5\times 10^{8}M_{\odot}$ and a GW frequency of $f_{\rm GW}=10^{-8.4}$ Hz. The remaining CGW parameters are fixed to $\psi=1.1$, $\cos\iota=-0.4$, and $\Phi_0=4.2$; these specific choices are not expected to significantly influence the localization performance. To test the robustness of parameter estimation against different signal properties, we additionally simulate two control configurations: a higher-mass system ($\mathcal{M}=1\times 10^{9}M_{\odot}$ at $10^{-8.4}$ Hz) and a higher-frequency system ($\mathcal{M}=5\times 10^{8}M_{\odot}$ at $10^{-8.1}$ Hz). We also simulate datasets at alternative sky locations to ensure our conclusions are not direction-dependent.

To isolate the impact of high-precision anchor pulsars from the geometrical benefits of a general array, we construct several specific PTA configurations. Our primary baseline is the 25-pulsar array. From this, we evaluate subsets containing only the 3 or 6 anchor pulsars, as well as reduced arrays of 19 and 22 pulsars that exclude these high-precision anchors.

\textit{4. Search Strategies.}
We adopt a Bayesian inference framework to recover the CGW parameters. For the anchor pulsars, we evaluate three distinct search strategies based on different distance prior assumptions:

\begin{itemize}
    \item \textbf{Ideal:} Perfect distance knowledge ($D_{\rm err} = 0$).
    \item \textbf{Sub-wavelength:} Distance uncertainty tightly constrained to $D_{\rm err}= 1{\rm pc}$, strictly below the GW wavelength ($\lambda_{\rm GW} \approx 2.4{\rm pc}$ for our fiducial signal).
    \item \textbf{Standard:} Current typical distance uncertainty, modeled as a 20\% fractional error.
\end{itemize}

For all non-anchor pulsars in the array, we apply the standard 20\% fractional distance uncertainty regardless of the overarching search strategy. We note that while a 20\% fractional uncertainty may underestimate the true distance errors, adopting larger uncertainties has no impact on the source localization. Once the distance error significantly exceeds wavelength $D_{\rm err} \gg \lambda_{GW}$, the phase of the pulsar term is already completely scrambled, rendering the exact magnitude of the prior error irrelevant. The pulsar distance and associated Earth-pulsar phase difference are parameterized as:
\begin{equation}
D_{p} = D_{p,0} + D_{\rm err} \cdot D_{\rm prior},
\end{equation}
\begin{equation}
\Delta\Phi = \Phi_{e} - \Phi_{p} + \Phi_{\rm prior},
\end{equation}
where $D_{p,0}$ denotes the nominal distance to the pulsar, and $D_{\rm prior}$ and $\Phi_{\rm prior}$ are stochastic prior variables.
In the Sub-wavelength and Standard searches, $D_{\rm prior}$ is drawn from a standard normal distribution $\mathcal{N}(0,1)$. The Standard search additionally assigns a uniform prior to the phase $\Phi_{\rm prior}$ to account for phase scrambling, whereas the coherent searches fix this offset to zero. Uniform priors are adopted over physically allowed ranges for all other intrinsic and extrinsic CGW parameters.

To determine the transition from the incoherent to the coherent localization regime, we also conduct searches varying the anchor distance uncertainties continuously from 0.2 pc up to 10.0 pc.
We simulated the pulsar timing data using the \texttt{libstempo}\ucite{libstempo} package. The simulated data were then analyzed using the \texttt{ENTERPRISE}\ucite{enterprise} and \texttt{ENTERPRISE\_EXTENSIONS}\ucite{enterprise_ext} software package for Bayesian inference, together with \texttt{PTMCMCSampler}\ucite{PTMCMCSampler_2017,Vousden2016} for stochastic sampling of the posterior distributions.

\textbf{Results.} 

\begin{figure}[htbp]
    \centering
    \includegraphics[width=\linewidth]{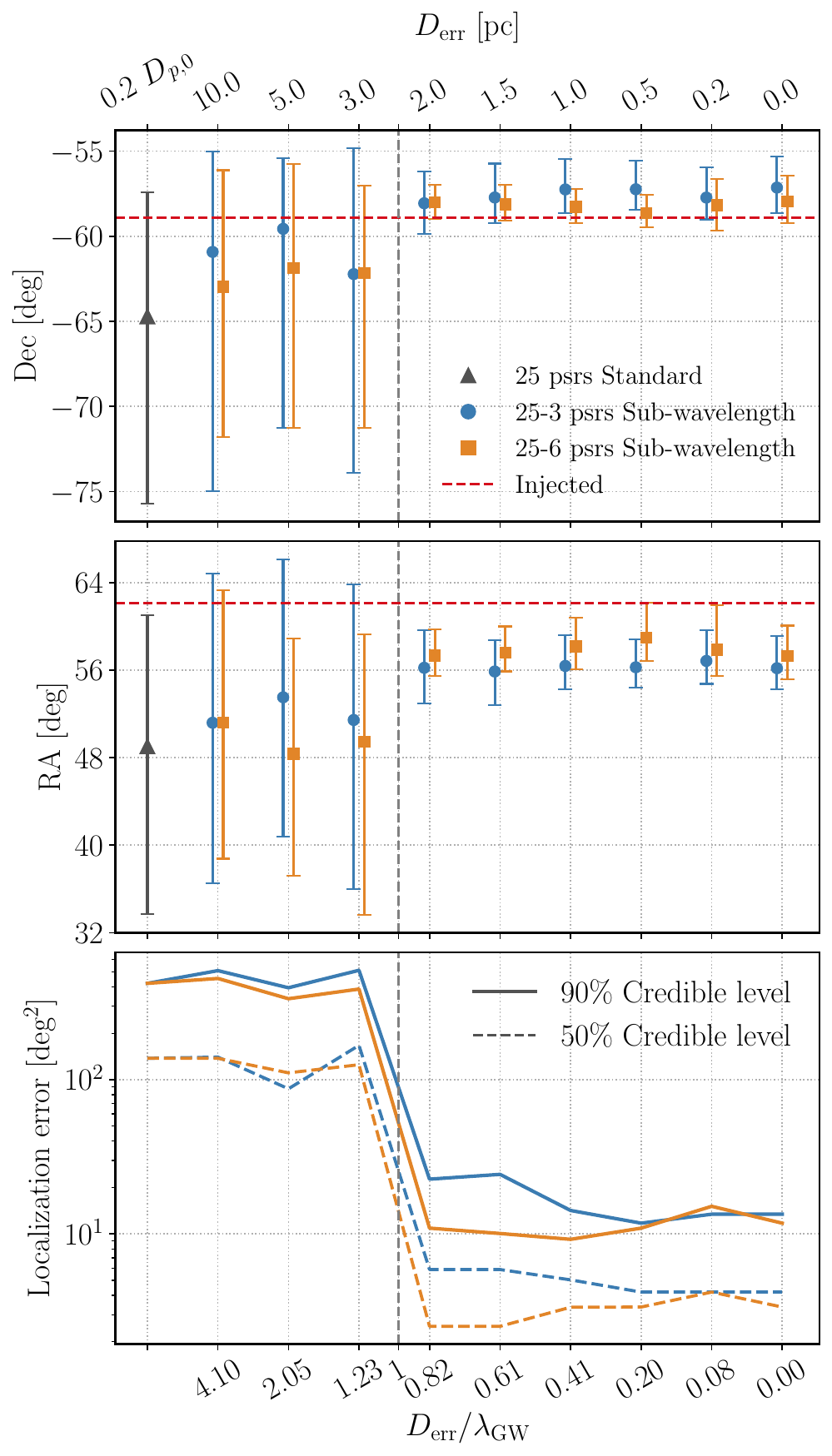}
    \caption{
    Dependence of sky localization precision on the ratio between pulsar distance uncertainty and the gravitational wavelength. 
    The $68\%$ credible intervals of RA and Dec (upper and middle panel) are shown for the 25-pulsar Standard search and for the sub-wavelength searches with three and six high-precision pulsars (25--3 and 25--6). The 90\% and 50\% credible level 2-D localization error is shown in the lower panel.
    }
\label{fig:25-3_25-6_derr_grad_with25_3_fixed}
\end{figure}

\textit{1. The Threshold Effect of Pulsar Distance Precision}

To determine the minimum astrometric precision required to meaningfully improve CGW source localization, we systematically evaluate how localization accuracy depends on the ratio between the pulsar distance uncertainty and the gravitational wavelength ($D_{\rm err}/\lambda_{\rm GW}$).
Figure \ref{fig:25-3_25-6_derr_grad_with25_3_fixed} presents the 68\% credible intervals for Declination (Dec) and Right Ascension (RA) across varying levels of distance precision. For this experiment, we utilize the 25-pulsar array at an $S/N$ of 20, assuming either three or six ``anchor" pulsars achieve specific distance uncertainties (ranging from 0.2 pc to 10.0 pc, while the remaining pulsars retain the standard 20\% fractional distance error.

\begin{table}[htbp]
\centering
\setlength{\tabcolsep}{4pt}
\begin{tabular}{lccc}
\hline
Array & Source & 90\% error (deg$^2$) & 50\% error (deg$^2$) \\ 
\hline
3-3 & Max & 1946.32 & 563.17 \\ 
6-6 & Max & 65.46 & 15.11 \\
19-0 & Max & 579.11 & 166.18 \\
22-0 & Max & 570.72 & 177.09 \\
25-0 & Max & 421.33 & 137.64 \\
25-3 & Max & 14.26 & 5.04 \\
25-6 & Max & 9.23 & 3.35 \\
25-6 & Min & 0.13 & 0.09 \\
25-6 & f81 & 0.099 & 0.019 \\
25-6 & M19 & 11.75 & 3.36 \\
25-25 & Max & 0.058 & 0.0048 \\
\hline
\end{tabular}
\caption{Sky localization uncertainties of CGWs for different array configurations. In the ``Array" column, the left number indicates the total number of pulsars, while the right number is the number of pulsars with sub-wavelength distance precision (except the last row with zero distance uncertainty assumed). In the ``source" column, Max and Min denote the sky location with maximum and minimum detection sensitivity for the given 25-pulsar array, respectively; ``f81" or ``M19" is the same as Max except that we adopt a higher GW frequency $f_{\rm GW}=10^{-8.1}$ Hz or higher chirp mass $\mathcal{M}=1\times10^{9}M_{\odot}$. In the last two columns, we list the 90\% and 50\% credibility localization errors. In all cases, the signal to noise ratio is 20.}
\label{tab:distance_configs}
\end{table}

\begin{figure*}[htbp]
    \centering
    \includegraphics[width=\linewidth]{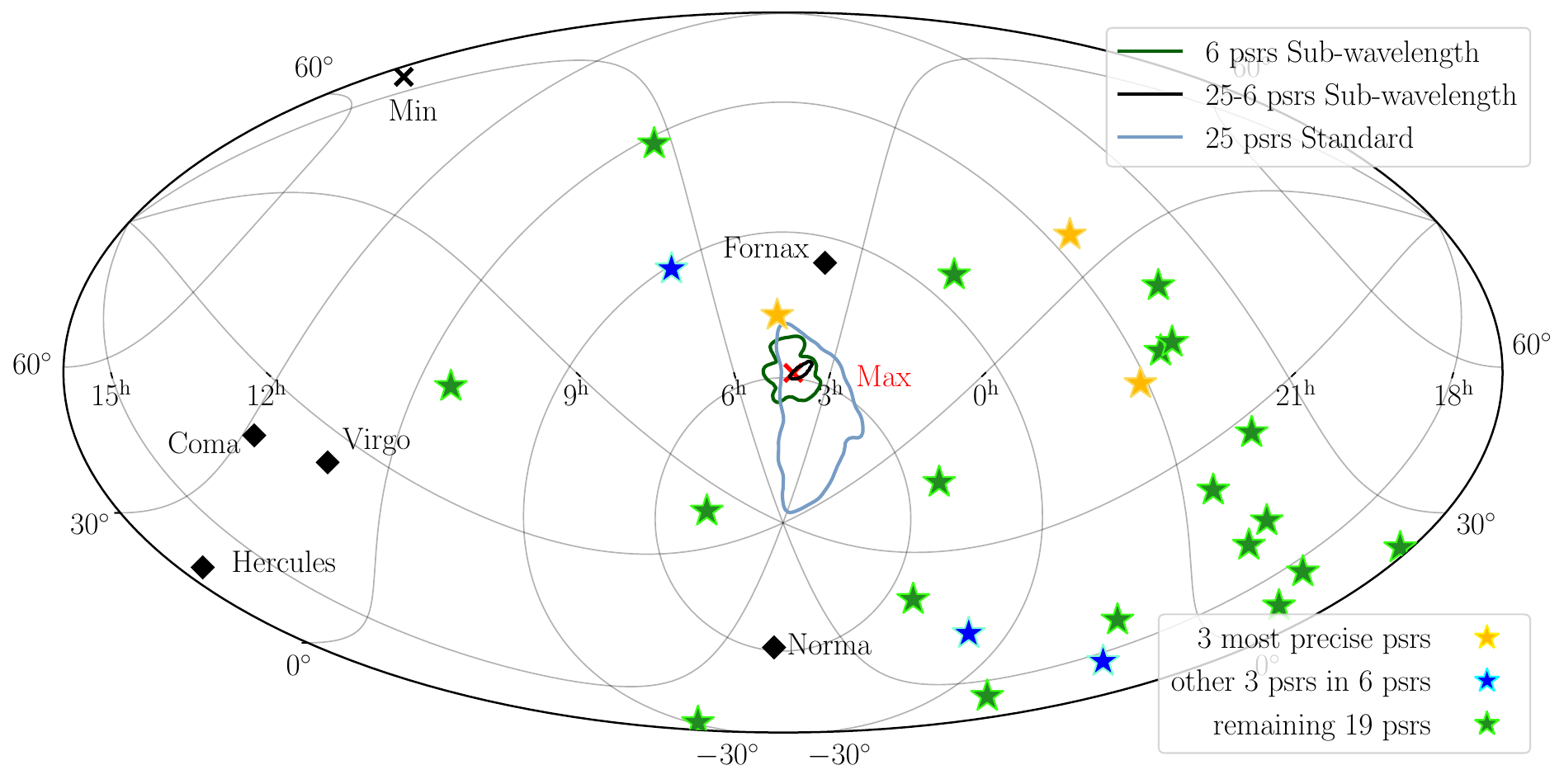}
    \caption{For different PTA configurations, we show the
    90\% credibility sky localization regions at $S/N=20$ for our reference CGW signal described in the main text. 
    The source is injected at the place marked by a red cross (labeled as ``Max"). Also marked in this sky map are 25 pulsars, 5 galaxy clusters, and the sky location of minimum detection sensitivity (labeled ``Min").
    }
\label{fig:skymap_center_90cl_legend_left}
\end{figure*}

\begin{figure}[htbp]
    \centering
    \includegraphics[width=\linewidth]{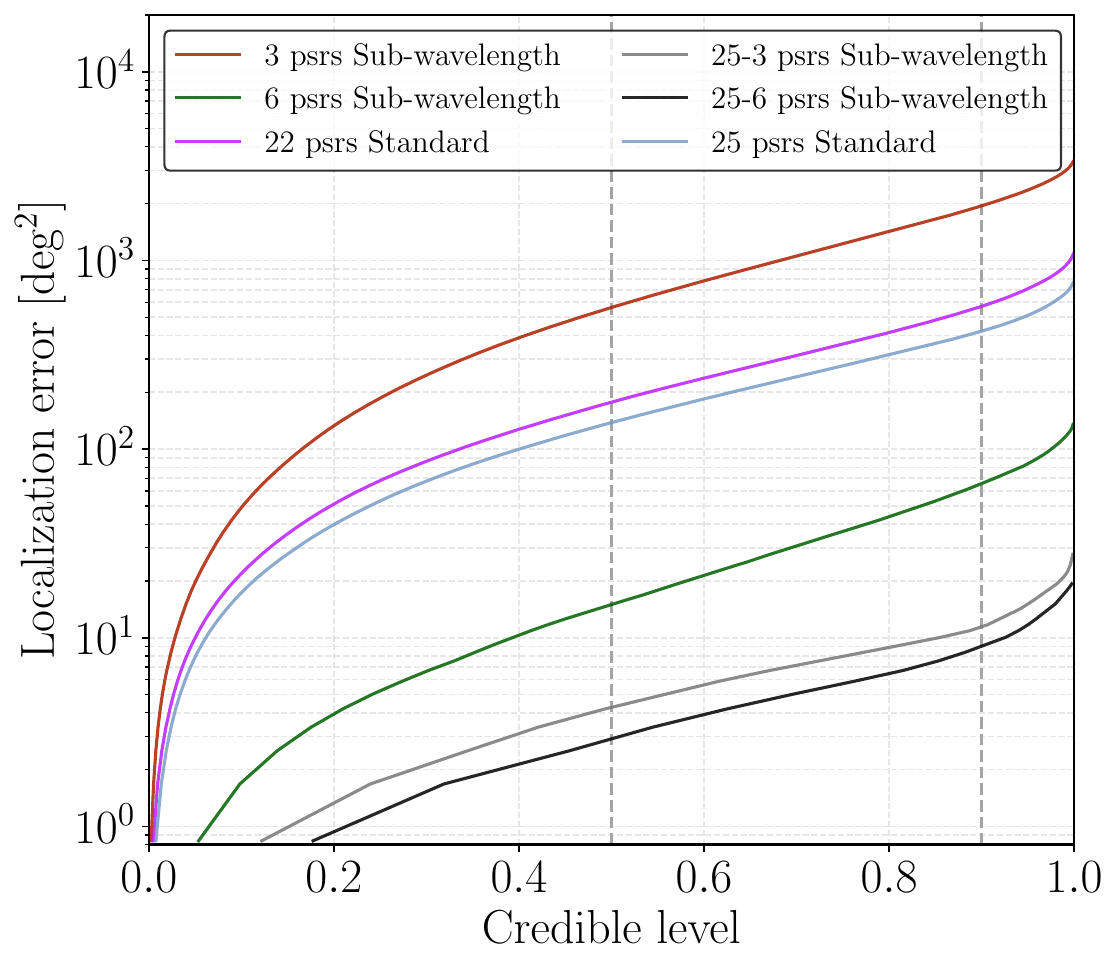}
    \caption{
    Sky localization capabilities of different PTA configurations for an injected CGW at the reference sky location and $S/N=20$ (see the main text and Table \ref{tab:distance_configs}).
    }
\label{fig:3_6_19_22_25_line}
\end{figure}

A clear threshold-like transition is observed. When the distance uncertainty of the anchor pulsars exceeds the gravitational wavelength ($D_{\rm err} > \lambda_{\rm GW}$), the localization capability remains relatively weak, offering only marginal improvements over the Standard search where all pulsars have low-precision measurements. However, once the distance uncertainty is constrained below the wavelength ($D_{\rm err} < \lambda_{\rm GW}$), the array transitions into a coherent regime. The credible sky region shrinks rapidly, indicating that the phase of the pulsar term has been successfully locked.

Interestingly, further improving the distance precision deep within this sub-wavelength regime (e.g., reducing $D_{\rm err}$ from 1.5 pc to 0.2 pc) yields diminishing returns.
Once the $D_{\rm err} < \lambda_{\rm GW}$ threshold is crossed, the array successfully transitions into a fully coherent regime. At this point, the sky localization area is no longer limited by the pulsar-term phase prior, but rather becomes dominated by the finite signal-to-noise ratio and the geometric distribution of the pulsars in the array.
We also note a slight deviation between the inferred Right Ascension (RA) and the injected value in Figure \ref{fig:25-3_25-6_derr_grad_with25_3_fixed}. This marginal offset arises from random noise realization, which shifts the maximum likelihood peak slightly away from the true injected parameters.

\textit{2. Relative Contribution of Anchor Pulsars vs. Array Size}

Having established that sub-wavelength anchor pulsars unlock high-precision localization within a comprehensive 25-pulsar network, we now explore the interplay between astrometric precision and array size. Specifically, we must isolate whether this enhanced localization is driven entirely by the high-precision anchors themselves, or if the geometrical baseline of the broader, standard-precision array remains an essential component.
Figure \ref{fig:skymap_center_90cl_legend_left} illustrates the 2D posterior sky localization regions for several distinct PTA configurations, Figure \ref{fig:3_6_19_22_25_line}  quantifies the corresponding sky localization area as a function of the credible level, and Table \ref{tab:distance_configs} summarizes the precise 90\% and 50\% localization errors.

A distinct hierarchical pattern emerges. Small arrays composed entirely of high-precision pulsars yield weaker localization constraints (e.g., 1946 deg$^2$ for the 3-3 array at the 90\% credible level) compared to larger arrays containing exclusively standard-precision pulsars (e.g., 571 deg$^2$ for the 22-0 array).
However, notably, the 6-6 array provides a remarkably tighter constraint ($65\text{ deg}^2$) than the 25-0 array. Despite having significantly fewer spatial baselines, the 6-6 array benefits from fully coherent pulsar terms, proving that securing phase coherence across just a half-dozen pulsars is vastly superior to incoherent addition across a large network.

When a subset of high-precision anchor pulsars is embedded within a large standard-precision array, the localization capability improves drastically. As seen in Figure \ref{fig:skymap_center_90cl_legend_left}, the posterior sky region for the 25-pulsar Standard array is poorly constrained due to unconstrained phase degeneracies. The introduction of six sub-wavelength anchors to form the mixed 25-pulsar array breaks these degeneracies, shrinking the 90\% credible sky area from 421.33 $\text{deg}^2$ in the Standard scenario (25-0) to just 9.23 $\text{deg}^2$.
To establish an absolute theoretical baseline for these improvements, we also calculated the localization error for a perfect 25-pulsar array where all pulsar distances are perfectly known (the 25-25 configuration).

As shown in the final row of Table \ref{tab:distance_configs}, this idealized array achieves an extraordinary 90\% credible sky area of just 0.058 $\text{deg}^2$ for the `Max' reference direction. Comparing this to our practical 25-6 array, which yields $9.23 \text{ deg}^2$ for the same `Max' direction, it is clear that expanding sub-wavelength precision to the entire array would tighten the constraints further by leveraging more coherent baselines. However, because securing sub-parsec astrometry for 25 pulsars is observationally prohibitive, the 25-6 configuration represents a realistic balance between observational feasibility and high-precision localization.
Table \ref{tab:distance_configs} also lists the localization errors for alternative signal parameters and sky locations, further confirming the broad applicability of our approach.

\textit{3. Robustness Against Signal Parameters}

\begin{figure*}[htbp]
    \centering
    \includegraphics[width=0.8\linewidth]{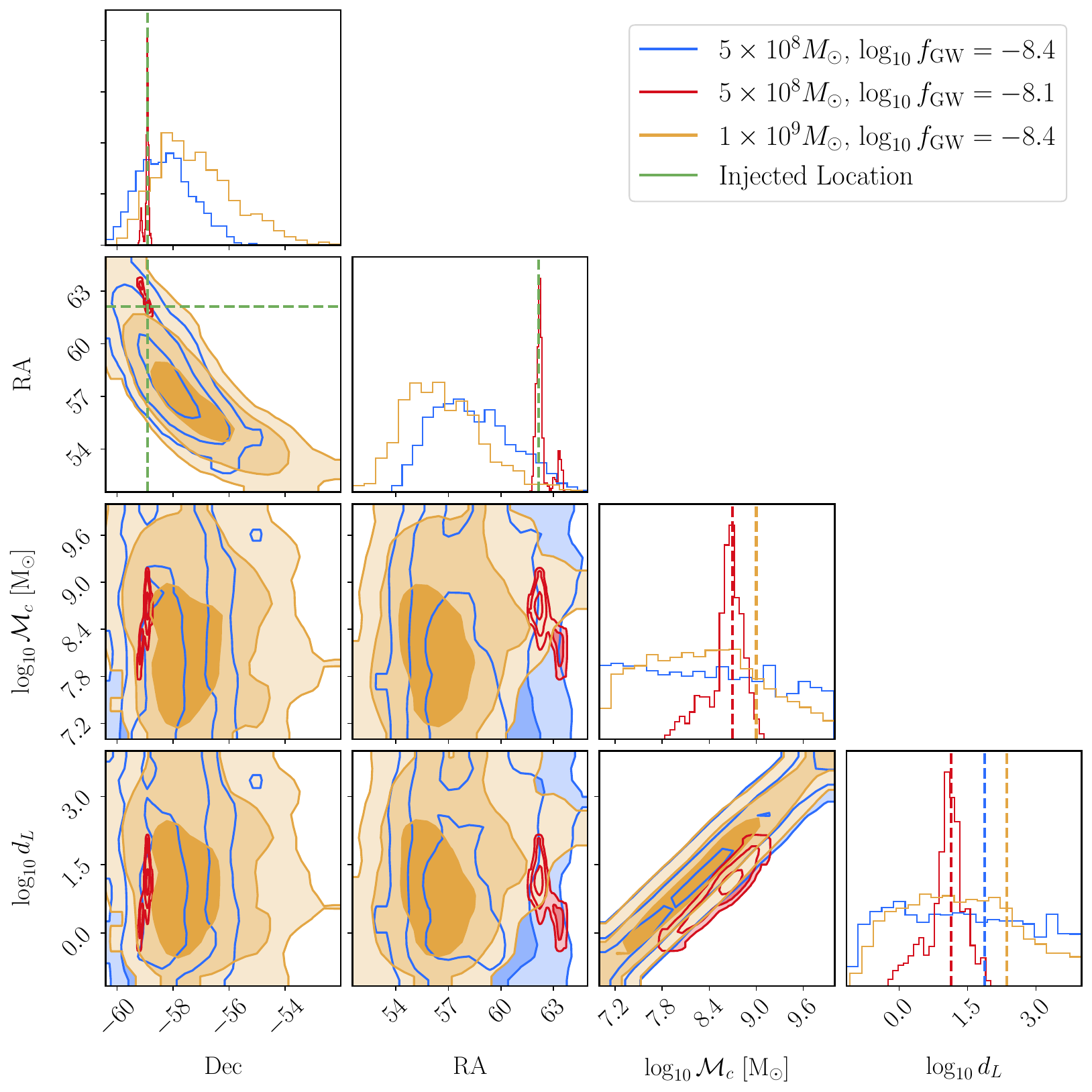}
    \caption{
    Posterior distributions of the key signal parameters obtained from the Bayesian search for three simulated signals. 
    All injections share the same sky location as the reference case used in the main text and have a signal-to-noise ratio of $S/N=20$. 
    The blue, red, and orange curves correspond to signals with different chirp mass or GW frequency.
    }
\label{fig:19_81_ra_dec}
\end{figure*}

To ensure our findings are not highly specific to the fiducial signal, we examine parameter recovery across different intrinsic source properties. Utilizing the mixed 25-pulsar array with six sub-wavelength anchors, we inject two additional CGW signals: one with a higher chirp mass ($\mathcal{M}=1\times10^{9}M_{\odot}$) and one with a higher GW frequency ($f_{\rm GW}=10^{-8.1}$ Hz). Table \ref{tab:distance_configs} summarizes the corresponding quantitative localization errors (labeled as ``M19" and ``f81", respectively).

Figure \ref{fig:19_81_ra_dec}  displays the posterior distributions for the key parameters of these two signals along with our fiducial signal. The localization performance for the higher-mass signal is comparable to the baseline, yielding a 90\% credible sky area of $11.75~\text{deg}^2$. In stark contrast, the higher-frequency signal yields a dramatic improvement in localization. The 90\% credible region shrinks by over two orders of magnitude to an extraordinary $0.099~\text{deg}^2$, alongside exceedingly tight constraints on luminosity distance and chirp mass.
While one might intuitively suspect that enhanced frequency evolution ($\dot{f}$) at higher frequencies breaks the parameter degeneracies, we verified that for these specific source parameters, the frequency evolution between the Earth and pulsar terms remains well below the frequency resolution of the 20-year dataset. Therefore, this striking improvement is driven purely by geometry: at higher frequencies, the gravitational wavelength is shorter. Consequently, the geometric phase delay across the array varies much more rapidly with small changes in the source's sky coordinates\ucite{taylor2026angular}, drastically enhancing the array's spatial resolving power.
However, we note a practical observational caveat: because the gravitational wavelength scales inversely with frequency, achieving the necessary sub-wavelength distance threshold ($D_{\rm err} < \lambda_{\rm GW}$) becomes significantly more demanding for higher-frequency sources.

\textit{4. Robustness Against Sky Locations}

To verify that the drastic improvement afforded by sub-wavelength anchor pulsars is applicable to realistic astrophysical targets, we analyze parameter recovery for potential SMBBHs located in the directions of five well-known galaxy clusters: Coma, Fornax, Hercules, Norma, and Virgo.

Table \ref{tab:Sky_locations} presents the 90\% and 50\% sky localization errors for these cluster directions using the 25-pulsar array under different search strategies. Across all investigated sky locations, the results are consistent. The Standard search (25-0) consistently produces large localization contours, ranging from 62.95 $\text{deg}^2$ for Coma up to 384.40 $\text{deg}^2$ for Hercules. The incorporation of sub-wavelength anchor pulsars successfully breaks the geometric degeneracies associated with these distinct lines of sight. With just three anchors (25-3), the 90\% credible area shrinks significantly. With six anchors (25-6), the array achieves exceptional precision—reducing the area to 0.60 $\text{deg}^2$ for Coma, 4.20 $\text{deg}^2$ for Hercules, and 1.68 $\text{deg}^2$ for Virgo. This demonstrates that the anchor pulsar strategy provides tight and reliable constraints for targeted multi-messenger follow-ups across a variety of realistic host environments.

\begin{table}[htbp]
\centering
\small
\setlength{\tabcolsep}{4pt} 
\begin{tabular}{lccc}
\hline
Array & Source & 90\% error (deg$^2$) & 50\% error (deg$^2$) \\
\hline
25-0 & Coma & 62.95 & 18.46 \\
25-3 & Coma & 29.37 & 8.39 \\
25-6 & Coma & 0.60 & 0.19 \\
25-0 & Fornax & 276.97 & 79.73 \\
25-3 & Fornax & 11.75 & 4.20 \\
25-6 & Fornax & 7.55 & 2.51 \\
25-0 & Hercules & 384.40 & 116.66 \\
25-3 & Hercules & 276.97 & 77.22 \\
25-6 & Hercules & 4.20 & 1.68 \\
25-0 & Norma & 175.41 & 53.71 \\
25-3 & Norma & 17.63 & 6.71 \\
25-6 & Norma & 0.12 & 0.041 \\
25-0 & Virgo & 243.40 & 71.34 \\
25-3 & Virgo & 108.26 & 32.73 \\
25-6 & Virgo & 1.68 & 0.84 \\
\hline
\end{tabular}
\caption{As in Table~\ref{tab:distance_configs}, but for potential SMBBHs in five galaxy clusters.}
\label{tab:Sky_locations}
\end{table}

\textbf{Discussion}
The dramatically improved sky localization has profound consequences for multi-messenger astronomy. In our realistic PTA simulations, satisfying the sub-wavelength condition shrinks the credible sky area to just a few square degrees. A localization region of this size ($\sim 0.1 - 7.6~\text{deg}^2$) is qualitatively different from the hundreds or thousands of square degrees yielded by standard searches; it makes cross-matching with existing large-scale galaxy surveys viable. This precision enables targeted galaxy catalog searches and practical wide-field electromagnetic imaging campaigns, bringing the identification of the SMBBH host galaxy and its subsequent astrophysical characterization firmly within observational reach.

To effectively phase-lock the array, an anchor pulsar must possess not only sub-wavelength astrometric precision, but also a strong response to the CGW signal.
As detailed in Appendix Table \ref{tab:pulsar_params}, fractional $S/N$ contributions vary significantly across the array. We demonstrated this dual requirement through a sanity check: applying the sub-wavelength prior strictly to the six pulsars with the \textit{lowest} $S/N$ contributions drastically degraded the localization performance to 396.99 $\text{deg}^2$ and 122.54 $\text{deg}^2$ at the 90\% and 50\% credible levels, respectively. Compared to the tight constraints of our default 25-6 configuration, this confirms that astrometric precision alone is insufficient without strong individual signal contributions.

From an observational standpoint, however, we cannot arbitrarily apply sub-wavelength distance priors to whichever pulsars happen to be most sensitive to a given signal. Observational reality restricts us to a small subset of nearby bright millisecond pulsars that are actually capable of reaching sub-parsec distance precision.
Fortunately, Table \ref{tab:pulsar_params} demonstrates that our observationally viable six-anchor subset contributes significantly to the network $S/N$ across diverse source directions. Because these anchors capture a substantial fraction of the signal strength, the 25-6 array consistently achieves high-precision localizations across all tested sky locations (Table \ref{tab:Sky_locations}).

For this given array configuration, we empirically find that a network $S/N \approx 20$ serves as a practical threshold for high-precision localization. While a lower network $S/N$ (e.g., $S/N=8$) can yield precise localization for certain source directions, we find that it falls short when applied to a diverse set of source locations.
To guarantee high-precision localization across a realistic, all-sky range of directions—such as our five galaxy cluster targets—we adopt a higher network $S/N \approx 20$ as the practical baseline threshold throughout our analysis. At this level, the anchor pulsars are virtually guaranteed to accumulate enough individual signal weight to effectively phase-lock the array and drive the localization area down to the scale of a few square degrees.

We note two recent studies that are complementary to our work. Grunthal et al. \cite{grunthal2026role} demonstrated that for high-frequency ($> \unit[10]{nHz}$) CGWs, leveraging pulsars at large distances (e.g., $\sim 4$ kpc) causes the Earth-term and pulsar-term frequencies to separate, passively mitigating pulsar-term self-noise in Earth-term-only searches. Analytical work by Taylor \cite{taylor2026angular} demonstrated that securing sub-wavelength distance precision actively unlocks Earth-pulsar interference, exponentially improving localization until the network hits its theoretical diffraction limit. Our work points out a cost-effective observational directive: rather than relying on pulsar distances to safely decouple and ignore the pulsar term, or demanding unfeasible array-wide sub-wavelength astrometry to reach theoretical limits, we demonstrate that coherently phase-locking a minimal core of just 3 to 6 anchor pulsars is practically sufficient to achieve high-precision source localization.

While capping the anchor subset at six pulsars represents a realistic observational baseline, it is instructive to consider the ultimate theoretical limits for next-generation facilities. If long-term advancements in pulsar astrometry eventually allow sub-parsec distance measurements for more pulsars, the PTA localization capabilities will continue to scale dynamically with the number of coherent baselines. Quantitatively, as demonstrated in Table \ref{tab:distance_configs}, expanding the sub-wavelength precision from 6 anchors to the entire 25-pulsar network shrinks the 90\% credible sky area for our reference `Max' direction from 9.23 $\text{deg}^2$ down to an extraordinary 0.058 $\text{deg}^2$. Therefore, while a core set of 3 to 6 anchors provides the critical geometric leap necessary to enable initial host-galaxy identification, continuing to expand the anchor sub-array toward 10 or more pulsars remains a highly valuable, long-term observational goal.
\medskip

\textbf{Conclusions.}

In this work, we quantitatively evaluated the impact of high-precision pulsar distance measurements on the sky localization of CGWs using realistic 20-year PTA datasets. Rather than demanding idealized astrometric precision across an entire network, we explored an observationally feasible strategy utilizing a minimal subset of high-precision ``anchor" pulsars embedded within a standard 25-pulsar array.

Our Bayesian parameter estimation reveals a critical, physical threshold effect. Once the distance uncertainty of anchor pulsars is reduced below the gravitational wavelength ($D_{\rm err}<\lambda_{\rm GW}$), the array seamlessly transitions into a fully coherent regime. This phase-locking completely breaks the severe geometric degeneracies that plague standard PTA searches, shrinking the 90\% credible sky localization area by nearly two orders of magnitude. We demonstrated that this anchor-driven enhancement is remarkably robust across varying intrinsic source parameters—such as chirp mass and GW frequency—as well as across different locations on the celestial sphere. Furthermore, improving astrometric precision deep into this sub-wavelength regime yields diminishing returns, establishing $D_{\rm err}<\lambda_{\rm GW}$ as the definitive target for future astrometric campaigns.

These findings chart a highly practical and cost-effective pathway toward precision multi-messenger astronomy with PTAs. By prioritizing intensive astrometric campaigns—such as high-precision VLBI and forthcoming SKA observations—for a small core of stable millisecond pulsars, the PTA community can achieve the high sky localization accuracy required to identify the electromagnetic counterparts and host galaxies of SMBBHs. Crossing this sub-wavelength threshold represents a transformative milestone, bridging the gap between nanohertz GW detection and comprehensive astrophysical characterization.
\medskip

\textit{Acknowledgements.}
This work is supported by the National Key Research and Development Program of China (No. 2023YFC2206704), the Fundamental Research Funds for the Central Universities, and the Supplemental Funds for Major Scientific Research Projects of Beijing Normal University (Zhuhai) under Project ZHPT2025001.


\medskip


\onecolumn

\begin{center}

{\large\bfseries Appendix A: Pulsar distances and noise parameters}
\end{center}

Table~\ref{tab:pulsar_params} summarizes the pulsar distances, the corresponding red- and 
white-noise parameters used in this work, and the signal-to-noise ratio of each pulsar for different CGW sources investigated in Tables \ref{tab:distance_configs} and \ref{tab:Sky_locations}. Here, we describe the current distance measurement precision for the six anchor pulsars considered in this work in more detail (all distances in kpc, and quoted uncertainties are at the 68\% credible level):
$0.329\pm 0.005$ for J0030$+$0451 \cite{Ding2023}, $0.1570\pm 0.0001$ for J0437$-$4715 \cite{reardon2024apj}, $0.26^{+0.07}_{-0.04}$ for J0636$-$3044 \cite{shamohammadi2024meerkat}, $0.388\pm 0.005$ for J1744$-$1134 \cite{antoniadis2023second1}, $1.158\pm 0.003$ for J1909$-$3744 \cite{liu2020}, and $0.269\pm 0.001$ for J2222$-$0137 \cite{ding2024}.
Note that kinematic distances (derived assuming General Relativity is correct) are quoted for PSR~J0437$-$4715 and J1909$-$3744 while other distances are calculated from parallax measurements.

\begin{table}[H]
\centering
\scriptsize
\setlength{\tabcolsep}{3pt}
\resizebox{\textwidth}{!}{%
\begin{tabular}{c c c c c c| c c c c c c c c c}
\hline
PSR & $D_{p,0}$ & EFAC & $\log_{10}$EQUAD & $\log_{10}A$ & $\gamma$ & Max & Min & f81 & M19 & Coma & Fornax & Hercules & Norma & Virgo \\
\hline

J0613$-$0200 & 1.010 & 0.83 & $-$8.74 & $\cdots$ & $\cdots$ & 5.73 & 51.17 & 17.67 & 6.40 & 13.78 & 5.90 & 0.19 & 1.73 & 5.65 \\
J1024$-$0719 & 1.080 & 0.77 & $-$7.33 & $\cdots$ & $\cdots$ & 3.87 & 22.34 & 9.78 & 0.52 & 26.01 & 3.70 & 7.01 & 5.59 & 8.00 \\
J1600$-$3053 & 1.390 & 0.65 & $-$6.83 & $-$13.86 & 2.92 & 0.10 & 1.12 & 4.96 & 0.60 & 4.00 & 0.25 & 1.73 & 2.12 & 2.39 \\
J1730$-$2304 & 0.510 & 0.89 & $-$7.09 & $\cdots$ & $\cdots$ & 1.47 & 2.02 & 4.42 & 2.50 & 20.17 & 1.37 & 6.58 & 5.01 & 13.93 \\
J1843$-$1113 & 1.997 & 0.75 & $-$8.20 & $\cdots$ & $\cdots$ & 1.37 & 10.37 & 8.07 & 1.35 & 13.23 & 1.70 & 11.66 & 10.00 & 10.94 \\
J1857+0943 & 1.150 & 0.97 & $-$7.99 & $\cdots$ & $\cdots$ & 1.04 & 25.09 & 3.05 & 1.11 & 9.86 & 0.67 & 4.40 & 1.11 & 8.47 \\
\hline
J0437$-$4715$^{\dagger}$ & 0.157 & 1.20 & $\cdots$ & $\cdots$ & $\cdots$ & 94.94 & 30.61 & 25.16 & 96.51 & 44.37 & 93.74 & 6.20 & 24.72 & 50.05 \\
J0636$-$3044\textsuperscript{*} & 0.260 & 1.04 & $\cdots$ & $\cdots$ & $\cdots$ & 0.26 & 7.76 & 0.78 & 0.26 & 1.58 & 1.23 & 0.09 & 1.58 & 1.12 \\
J1946$-$5403 & 1.300 & 0.97 & $\cdots$ & $\cdots$ & $\cdots$ & 3.75 & 2.64 & 6.57 & 2.92 & 4.22 & 3.85 & 1.83 & 6.59 & 2.69 \\
J2222$-$0137$^{\dagger}$ & 0.269 & 1.06 & $\cdots$ & $\cdots$ & $\cdots$ & 0.77 & 18.82 & 2.53 & 0.71 & 0.88 & 1.78 & 1.38 & 2.23 & 0.19 \\
J2241$-$5236 & 1.050 & 1.05 & $\cdots$ & $\cdots$ & $\cdots$ & 21.55 & 7.24 & 65.22 & 21.60 & 7.12 & 27.74 & 12.12 & 59.40 & 13.26 \\
\hline
J0030+0451$^{\dagger}$ & 0.329 & 1.05 & $-$7.83 & $-$14.20 & 4.01 & 0.56 & 1.13 & 4.01 & 0.55 & 0.03 & 0.19 & 0.17 & 0.47 & 0.07 \\
J1640+2224 & 1.400 & 1.07 & $-$7.84 & $-$14.71 & 0.94 & 0.20 & 35.11 & 1.58 & 0.07 & 12.63 & 0.06 & 6.96 & 4.32 & 5.70 \\
J1741+1351 & 1.800 & 1.01 & $-$7.42 & $-$15.65 & 4.83 & 0.81 & 10.91 & 2.93 & 0.84 & 9.00 & 0.24 & 7.92 & 7.38 & 7.79 \\
J1909$-$3744\textsuperscript{*} & 1.158 & 1.03 & $-$8.04 & $-$17.40 & 3.33 & 18.82 & 28.83 & 48.00 & 8.65 & 42.12 & 15.12 & 37.91 & 44.25 & 35.40 \\
J1911+1347 & 3.500 & 1.02 & $-$7.53 & $-$16.71 & 3.87 & 0.82 & 22.92 & 4.04 & 1.00 & 6.85 & 0.10 & 8.88 & 5.54 & 4.24 \\
J2017+0603 & 1.560 & 1.01 & $-$7.85 & $-$15.65 & 3.07 & 0.52 & 14.98 & 3.60 & 0.88 & 4.06 & 0.81 & 2.41 & 1.87 & 2.96 \\
J2043+1711 & 1.400 & 1.02 & $-$8.16 & $-$17.80 & 3.49 & 1.60 & 9.85 & 6.47 & 0.27 & 9.09 & 2.20 & 9.14 & 4.37 & 6.54 \\
J2234+0611 & 0.970 & 1.00 & $-$8.14 & $-$13.70 & 2.30 & 0.77 & 3.45 & 3.39 & 0.12 & 1.05 & 0.50 & 1.26 & 0.67 & 0.42 \\
J2234+0944 & 0.900 & 1.07 & $-$8.43 & $-$18.22 & 2.29 & 0.85 & 10.06 & 0.66 & 0.26 & 2.16 & 1.46 & 0.39 & 1.26 & 0.33 \\
J2317+1439 & 1.700 & 1.04 & $-$7.38 & $-$19.45 & 3.10 & 1.48 & 13.22 & 6.32 & 1.59 & 0.05 & 2.38 & 0.51 & 1.73 & 0.94 \\
\hline
J0125$-$2327 & 1.200 & 1.01 & $-$7.28 & $-$18.11 & 3.80 & 2.03 & 9.63 & 27.46 & 0.56 & 0.21 & 9.80 & 3.06 & 4.38 & 0.58 \\
J1017$-$7156 & 1.800 & 1.04 & $-$8.87 & $-$18.41 & 6.04 & 6.01 & 13.95 & 13.41 & 5.65 & 13.16 & 0.51 & 6.83 & 13.94 & 10.32 \\
J1713+0747 & 1.070 & 1.10 & $-$7.49 & $-$19.40 & 0.53 & 6.15 & 2.79 & 33.84 & 6.49 & 51.86 & 4.01 & 87.75 & 57.63 & 72.45 \\
J1744$-$1134\textsuperscript{*} & 0.388 & 1.00 & $-$7.39 & $-$17.56 & 5.50 & 3.88 & 33.39 & 1.84 & 2.21 & 37.00 & 3.84 & 9.15 & 4.52 & 7.20 \\
\hline

\end{tabular}%
}

\caption{
Noise parameters of 25 millisecond pulsars. $D_{p,0}$ is the pulsar distance in kpc, EFAC and $\log_{10}$EQUAD are white noise parameters, and $\log_{10}A$ and $\gamma$ are the red noise amplitude and spectral index. The rightmost nine columns show the signal to noise ratio of each pulsar for a signal with a total network $S/N = 100$ of the entire array; source names are the same as Tables \ref{tab:distance_configs} and \ref{tab:Sky_locations}. From top to bottom in each block of rows, the noise parameters are taken from public data sets of EPTA+InPTA, MPTA, NANOGrav, and PPTA, respectively. 
Pulsar names with $^{\dagger}$ denote the three-pulsar anchor subset adopted in this work, while $^*$ denotes the other three pulsars in the six-pulsar core subset.
}
\label{tab:pulsar_params}
\end{table}


\end{document}